\documentclass[12pt]{article}

\usepackage{newtxtext,newtxmath}

\usepackage{graphicx}
\usepackage{soul}

\usepackage[letterpaper,margin=1in]{geometry}

\renewenvironment{abstract}
	{\quotation}
	{\endquotation}

\date{\scalefont{0.7}\today}

\makeatletter
\renewcommand{\fnum@figure}{\textbf{Figure \thefigure}}
\renewcommand{\fnum@table}{\textbf{Table \thetable}}
\makeatother

\usepackage{scicite}

\usepackage{url}

\def\scititle{\scalefont{0.82}Caustic-Driven Fluidic Microlenses for Enhanced Nonlinear and High-Energy-Density Physics \\}
\title{\bfseries \boldmath \scititle}

\author{
	Sourabh Singh $^{1}$,
        S. Sree Harsha$^{1}$,
	Tamanna$^{1}$,
	Prashant~Kumar~Singh$^{1\ast}$\and
	\small$^{1}$Tata Institute of Fundamental Research Hyderabad, 36/P Gopanpally, Hyderabad 500046, Telangana, India \and
	\small$^\ast$Corresponding author. Email: pks@tifrh.res.in
}

\begin{document} 

\maketitle

\begin{abstract} \bfseries \boldmath
We demonstrate that caustic microlensing occurring in a liquid jet efficiently drives linear, nonlinear, and high–energy–density phenomena. In linear regime, caustics provide localized focusing, distinct from external high–NA optics. In nonlinear regime, they enhance the input field at liquid–air interface and boost surface-sensitive processes. In high-energy-density domain, caustic-driven localized laser absorption generates gigapascal shocks using microjoule femtosecond pulses, with scalability up to repetition rates up to 0.2 MHz. Our Caustic-driven fluidic microlensing offers opportunities for surface nonlinear optics, ultrafast and high-energy density science.

\end{abstract}

\noindent

\section*{}

Caustics, formed by the coalescence of multiple rays under reflection or refraction, manifest themselves as bright focusing features \cite{CausticsBook, BerryCautics1980}. Being intrinsic to all wave phenomena \cite{CausticsLeonardo}, caustics are widely observed in systems such as acoustics\cite{acousticBook}, water waves \cite{CauticsTsunami}, reflection from the ionosphere \cite{CausticsIonosphere}, radio-caustics in the interstellar medium \cite{RadioCaustics}, matter waves \cite{AtomLaser} and high-harmonic generation\cite{CausticsHHG1, CausticsHHG2}. Caustics occur across spatial scales, from subatomic phenomena such as nuclear scattering \cite{CausticsNuclear} and electron microscopy \cite{CausticsElectron}, to cosmic events such as gravitational microlensing \cite{CausticsGravity}. Due to their structurally stable nature against perturbations \cite{CausticsStable}, caustics offer a robust platform for microlensing. However, despite their ubiquity, the potential of caustics to control and concentrate optical energy in strong-field and ultrafast regimes remains largely unexplored. Generation of extreme light localization typically requires invoking high-numerical-aperture optics \cite{1997MazurAPL, HighNATeraPascal, HighNAVoid2006, VogelPRL2008, HighNASapphire}, dielectric microlensing \cite{1994PRLDroplet, 1986ScienceDroplet, 2009NatureLens, 2019NatComSuperlensing, 2022SciAdvMitrochondria, lipid_droplet}, or engineered optically polished surfaces \cite{2011NatComSuperlensing}, an approach that is alignment-sensitive and often unsuitable for high-repetition-rate operation. In contrast, caustic microlensing, naturally arising from refractive curved dielectric interfaces, can offer structurally stable envelopes of energy concentration without any external element \cite{2012autofocuscaustics}. Specifically, asymmetric illumination of a dielectric microcylinder can lead to the generation of curved photonic structures \cite{2020OptLetHook}, assisted by caustics\cite{CausticsBook}.

Here, we demonstrate that asymmetric irradiation of a flowing liquid microjet with femtosecond pulses generates fold-type optical caustics. This off-axis caustic microlensing configuration outperforms conventional on-axis geometries across regimes spanning linear to nonlinear optics and ultimately towards the creation of high-energy-density conditions. In the linear regime, the flowing liquid jet serves as a stable caustic lens, achieving one-dimensional tight focusing with modest, low–numerical-aperture input optics. We further show that this approach naturally localizes the highest intensity at the liquid-air interface, without any additional external requirement of precise alignment. We extend this microlensing mechanism into the nonlinear regime by exploiting the intrinsic localization of intensity at the liquid–air interface. This is supported by the observation of enhanced surface-sensitive nonlinear phenomena such as third-harmonic generation (THG) \cite{1995PRATHG}. Finally, we use this microlensing geometry to create localized high-energy-density conditions. This feature was witnessed by the formation of a microplasma sphere, followed by the generation of GPa-scale shock waves. In contrast to single-shot microexplosion experiments in solid dielectrics that rely on external high-NA optics \cite{HighNASapphire, HighNATeraPascal}, our approach uses caustic microlensing in a self-refreshing liquid jet to reach high-energy-density conditions at an unprecedented 0.2-MHz repetition rate. This demonstrates a scalable route toward megahertz-rate microexplosion experiments \cite{2018OptExpLiquidJet} driven by microjoule-level femtosecond oscillators \cite{2008NatPhotOscillator}.  Together, these natural microlensing-based platforms could advance areas such as optofluidics \cite{optofluidic}, surface-driven nonlinear optics \cite{1995PRATHG}, warm dense matter \cite{2008ScienceWDM}, shock-driven physics\cite{2021NatPhysWaterPhase}, and high-throughput ultrafast science \cite{2018NatComXFEL, 2021NatureAIdata}.

\section*{Results}

\subsection*{Linear regime: caustics microlensing in flowing liquid jet}

The experiments were performed at TIFR-Hyderabad using a Yb:KGW laser (Pharos, Light Conversion) delivering 170-fs pulses at 1034 nm, up to 200 kHz, with few $\mu$J pulse energy at focus. A 7-mm diameter beam was focused perpendicularly onto a 100–110 $\mu$m laminar liquid jet (deionized water) using an f = 100 mm gold-coated OAP mirror. Jets were generated by a 2-bar pressurized reservoir and interacted with the laser 2–3 mm below the nozzle to avoid Rayleigh–Plateau breakup. The basic concept of exciting caustics microlensing under asymmetric irradiation geometry, in a dielectric cylindrical jet and its comparison with symmetric lensing is shown in Fig. 1, (A and B). For an incident laser beam of radius 5$\times$ smaller than the radius of a cylindrical jet, the irradiation could be considered localized. In such scenarios, one could continuously tune the interaction geometry from the on-axis ($0^{\circ} \pm 8^{\circ}$), symmetric lensing (Fig. 1A), to the near grazing angle of  ($45^{\circ} \pm 8^{\circ}$) extreme asymmetric lensing (Fig. 1B). To directly observe the reshaping of incident laser spot due to dielectric microlensing, experiments were performed at an effective intensity well below the liquid breakdown threshold to avoid any degradation of beam profile due to plasma formation (Fig. S4). In the absence of a liquid jet, the measurement of the input reference laser focal spot shows a near cicular Gaussian shape, having a $\frac{1}{e^2}$ beam waist radius of $\omega_{0x}$ = 12.9 $\mu$m, $\omega_{0y}$ = 11.4 $\mu$m), Fig. 1C. For on-axis, symmetric microlensing, the input circular focal spot gets transformed into an elliptical shape ($\omega_{0x}$ = 4.1 $\mu$m, $\omega_{0y}$ = 11.2 $\mu$m) after traversing through the liquid jet due to cylindrical lensing acting only in $x$-direction, while leaving $y$-axis intact (Fig. 1D). During the off-centre, asymmetric geometry, a much tighter focusing is observed, where laser focal spot displays nearly a line focus ($\omega_{0x}$ = 1.5 $\mu$m, $\omega_{0y}$ = 11.3 $\mu$m), Fig. 1E. Additionally, the focal spot in the asymmetric geometry (Fig. 1E) displays one-sided low intensity wings, represented by finite-energy Airy beam pattern \cite{2007OLFiniteAiry}, a characteristic of fold catastrophe\cite{FoldCausticPRL, CausticsBook}, detailed in Supplementary Fig. S1 (C to F).


\begin{figure}[h!] 
	\centering
	\includegraphics[width=1.0\textwidth]{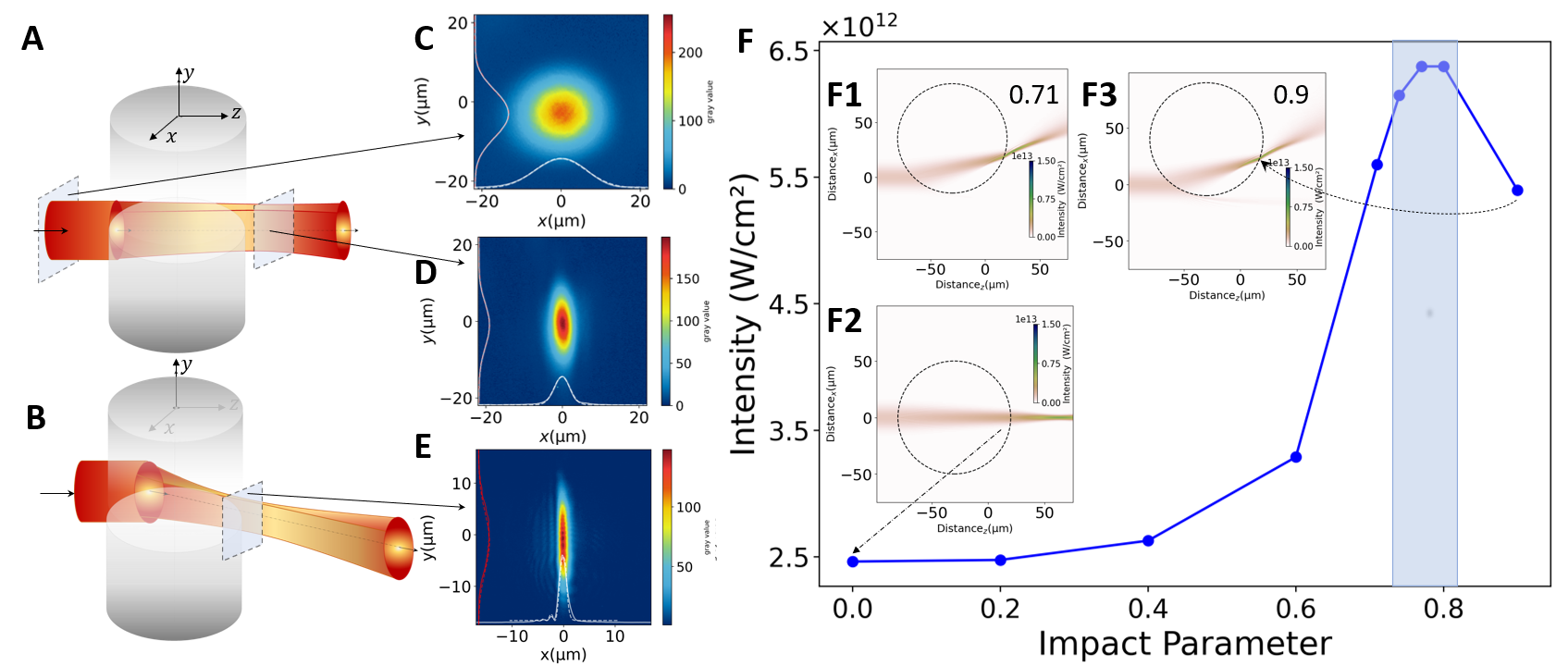} 
	\caption{\textbf{Symmetric and asymmetric Caustics microlensing in cylindrical liquid jet.}
Sketch of incident Gaussian laser pulse propagating along $z$ direction, undergoing lensing through a cylindrical liquid jet in symmetric (\textbf{A}) or asymmetric (\textbf{B}) geometry. (\textbf{C}), (\textbf{D}) and (\textbf{E}) are experimentally measured laser focal profiles, in scenarios corresponding to pre-liquid lensing, post-symmetric lensing and post-asymmetric lensing, respectively. (\textbf{F}) Numerical simulation demonstrates the variation of peak laser intensity, estimated at the rear surface of the cylindrical jet ($R$ = 50~\textmu m), for different impact parameters, considering incident laser of 170 femtosecond pulse duration, Gaussian focal spot (waist = 12.9 $\mu$m) and 1 $\mu$J energy. The insets (\textbf{F1}), (\textbf{F2}) and (\textbf{F3}) show the lensing occurring in the $x-z$ plane for impact parameters of 0, 0.71 and 0.9, respectively. The red arrow indicates the direction of the incident laser beam. }
	\label{fig:example} 
\end{figure}

The degree of asymmetry can be quantified in terms of the impact parameter, defined as $d/R$, where $d$ is the offset distance between the laser and jet centre, and $R$ is the jet radius. By this definition, the impact parameter of value 0 will correspond to the symmetric case, whereas the value of 1 will correspond to an extreme asymmetry with an incident laser beam grazing the cylinder edge. For a detailed understanding of impact parameter-dependent microlensing, we resort to numerical simulations, detailed in the Methods. As the cylindrical microlensing occurs only along one axis, simulations are performed in the 2-dimensional geometry ($x-z$ plane). For different values of impact parameter, the variation of laser intensity, estimated at the rear surface of the cylindrical jet, is presented in Fig. 1F. Moving from symmetrical geometry (impact parameter of 0) to a higher degree of asymmetry, the local intensity realized at the rear surface of jet, shows a steep enhancement. The intensity enhancement trend reaches its peak at 0.77, followed by a fall, as near the grazing geometry, part of the incident laser pulse starts missing the cylinder surface. The transition from symmetric to the asymmetric case is presented in Fig. 1, (F1 to F3), where lensing occurring in the $x-z$ plane for impact parameters of 0, 0.71 and 0.9, respectively, is shown. Basically, on increasing the impact parameter, the incident laser pulse of finite width progressively experiences a dielectric interface of increasing curvature \cite{1989AplOptDroplet}. This leads to stronger refractive lensing and brings the highest intensity spot closer to the jet rear surface. For an impact parameter of 0, the maximum laser intensity plane is located nearly 40 $\mu$m further from the jet rear surface, whereas for the asymmetric case (impact parameter $\ge$ 0.7), the peak intensity lies at the jet rear end. To some extent, the effect of large curvature near the grazing angle compensates for the partial loss that occurred as the laser beam misses intercepting the cylinder (Fig. 1F3).

\subsection*{Nonlinear regime: Caustics microlensing for interface sensitive phenomenon}
Beyond linear amplification of the incident laser intensity (Fig. 1F), caustic microlensing maintains the optimal focal plane at the jet’s rear surface (Fig. S2, S3) and thereby enhances nonlinear processes that occur mainly at the interface rather than in the bulk medium \cite{BoydBook}. To elucidate this mechanism, we performed numerical simulations tracking the full evolution of the laser intensity across the jet, capturing the influence of curvature before, during, and after cylindrical lensing in both symmetric and asymmetric configurations (Fig. 2A). These computations were performed considering only linear response of the dielectric refractive medium, without factoring in the non-linear propagation effects such as self-focusing or plasma effects, details in the Supplementary. Compared to the nearly flat reference input irradiance (red line) of a loosely focused beam with a large Rayleigh length ($>$ 400 $\mu$m), the symmetric lensing configuration (orange curve) gradually enhances the laser intensity along the propagation axis. This amplification continues beyond the rear surface of the liquid jet, with the intensity peaking approximately 40 $\mu$m further in the air (Fig. 2A). In contrast, for asymmetric geometry (Green Curve), despite the Fresnel loss at the first air-dielectric interface, the stronger focusing leads to nearly 5$\times$ enhancement of laser intensity at the rear edge of the liquid cylinder (Fig. 2A). The curvature-dependent stronger lensing is also manifested as a much narrower intensity distribution for the 0.71 impact parameter, than one for the impact parameter of 0.  Although the overall peak intensity received for symmetric and asymmetric cases is nearly the same, the key distinction lies in the position of the peak intensity for these two geometries. These offsets in the peak intensity locations can be drastically manifested in a physical phenomenon which is highly active at the dielectric-air interface. Third harmonic generation (THG) is one such process that, for an isotropic liquid, like water, has nearly zero bulk response, except at the interface where the bulk inversion symmetry is broken and THG produces a strong signal \cite{1995PRATHG}. To demonstrate the critical role of asymmetric lensing in enhancing the interface-sensitive nonlinear processes, we measured the THG signal generated at the water jet–air boundary. 

\begin{figure}[h!] 
	\centering
	\includegraphics[width=1.0\textwidth]{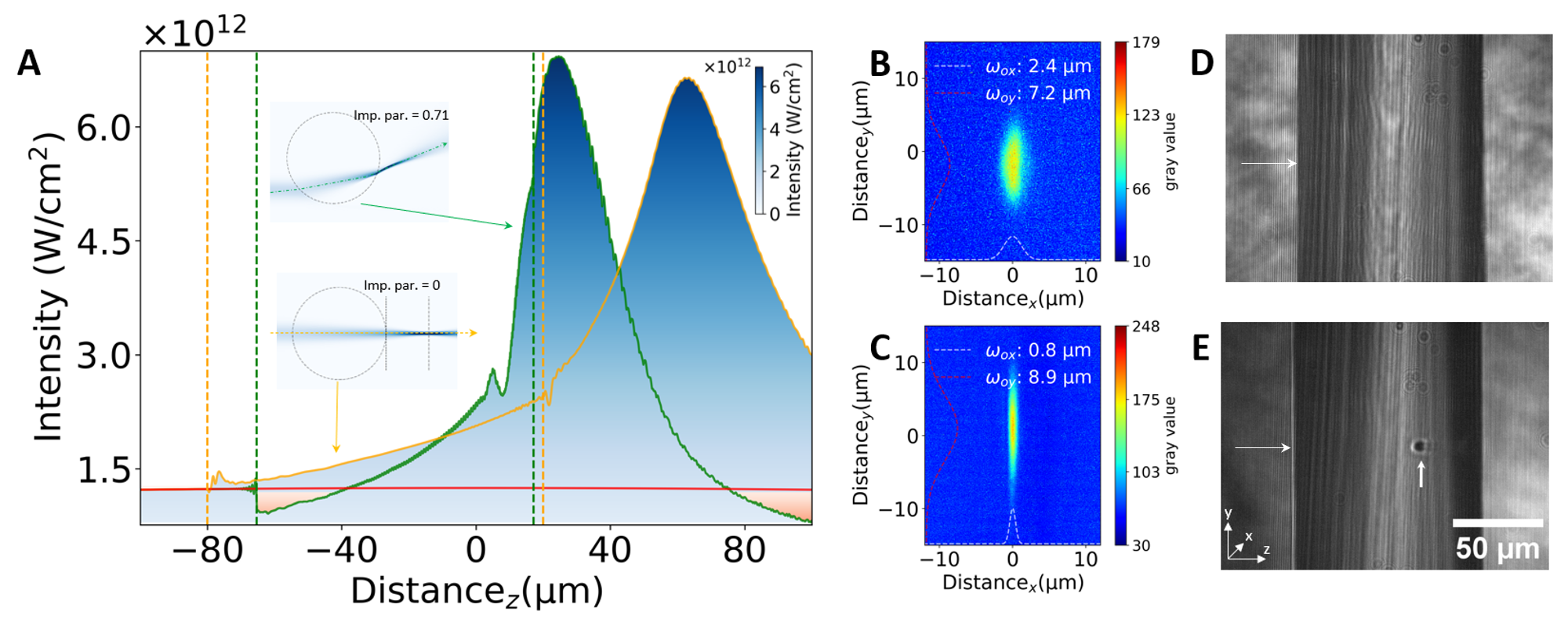} 
	\caption{\textbf{Caustics microlensing lensing to enhance interface phenomenon.}
(\textbf{A}) Simulated laser intensity variation during the entire course of lensing for impact parameters of 0 and 0.71, represented by orange and green curves, respectively. The horizontal solid red line represents reference laser intensity in the absence of a liquid jet. The vertical dashed lines represent the locations of cylinder entry and exit surfaces for symmetric (orange color) and asymmetric cases (green color).  Experimental measurements of the THG beam profiles recorded for a laser energy of 0.91 $\mu$J, at the jet rear surface for symmetric and asymmetric case, shown in (\textbf{B}) and (\textbf{C}) respectively. Experimental Shadowgram images of the liquid-jet irradiated by femtosecond laser pulse at 2.0 $\mu$J energy for impact parameter of 0 (\textbf{D}) and 0.71 (\textbf{E}), at the time-delay of 2.0 nanosecond. The horizontal white arrow in (D) and (E) indicates laser irradiation on the front surface, while the vertical white arrow in (E) shows the formation of a localized plasma channel. As the backlight probe beam is also subjected to the strong cylindrical lensing, in the shadowgram image (E), the plasma blob appears shifted inwards from the rear edge.}
\label{Shadow_Image0}
\end{figure}

Experimental measurements of the THG beam profiles recorded in the forward direction at the jet rear surface for symmetric and asymmetric cases are shown in Fig. 2B and 2C, respectively, detailed in Methods. Similar to the fundamental beam profiles (Fig. 1 D and E), the THG beam shapes are elliptical. However, due to third-order nonlinearity, the THG beam width for symmetric ($\omega_{x}^{THG}$ = 2.4 $\mu$m, $\omega_{y}^{THG}$ = 7.2 $\mu$m ) and asymmetric ($\omega_{x}^{THG}$ = 0.8 $\mu$m, $\omega_{y}^{THG}$ = 8.9 $\mu$m) are much smaller than their respective fundamental beams, Fig. 1 D and E. Furthermore, the caustics-driven asymmetric finite energy Airy wings present in the fundamental pulse (Fig. 1E), get drastically suppressed during the third-harmonic upconversion (Fig. 2C), due to non-linear intensity dependence filtering. As expected, due to higher laser intensity at the rear interface, the THG signal for asymmetric geometry is nearly 10 $\times$ higher than the symmetric case, as detailed in Fig. 4B.

At higher laser energies, the laser-matter interaction enters a regime where higher-order nonlinear processes beyond third-order effects become accessible \cite{BoydBook}. An infrared laser of 1.2 eV photon energy (wavelength of 1030nm), may require 6 or higher photons to drive multiphoton ionization (6.5 eV) or autoionization (9.5 eV) processes in water \cite{2015PRBBreakdown, 2016PRBBreakdown}. For laser parameters related to our experiment (1030nm, 170fs), the seed photoelectron via avalanche ionization can trigger the optical breakdown of the water at the peak intensity of $7\times 10^{12} W cm^{-2}$\cite{2016PRBBreakdown}. Towards this, space and time-resolved optical shadowgrams were recorded under symmetric (Fig. 2D) and asymmetric (Fig. 2E) irradiation geometries, with input laser energy of 2.0 $\mu$J. Consistent with the simulation results (Fig. 2A), at 2.0 $\mu$J laser energy, only the asymmetric geometry-driven focusing succeeds in surpassing the required intensity threshold for optical breakdown of the water jet. The steep variation of laser intensity inside the jet for the asymmetric case (Fig. 2A) also ensures that the breakdown region remains strongly localized, manifesting as a small plasma blob, Fig. 2E. The microlensing-driven plasma formation also turns out to be independent of the choice of horizontal or vertical polarization of the incident laser pulse, Fig. S3. These plasma structures, aided by nonlinear absorption, drive localized high-energy-density regions crucial for strong-field ultrafast studies.

\subsection*{Curvature driven nonlinear absorption and plasma formation}
Ultrafast laser–driven microlensing in transparent dielectrics provides an efficient route to achieve high-energy-density conditions in table-top experiments, as demonstrated in several earlier foundational work \cite{1997MazurAPL, HighNATeraPascal, HighNAVoid2006, VogelPRL2008, HighNASapphire, plasmawater2002, 2004AplOptDroplet, plasmawater2006, plasmawater2016, Dropshaping2015, plasmawater2018, plasmawater2020}. To understand how curvature-driven microlensing governs nonlinear energy deposition in transparent media, we investigated the laser–jet interaction by scaling up the input energies. As the incident laser intensity in the air gets clamped to the order of $10^{13} W cm^{-2}$\cite{2007Filamentation}, the laser absorption in a dielectric, transparent medium predominantly occurs via ionization and inverse bremsstrahlung mechanisms \cite{1999VogelWater}. Experiments were carried out to see the onset of nonlinear transmission and reflection from the water jet by varying input laser energy (Fig. 3 A and A1, detailed in Methods). At lower irradiance energy ($<$ 1 $\mu$J), the transmission data shows no sign of energy depletion. However, on a further increase of energy, at first, asymmetric geometry shows a drop of transmission at the threshold value of  1.1 $\mu$J, followed by the symmetric case at 1.5 $\mu$J (Fig. 3A). Apart from thresholding behaviour, the transmission data, does not show any drastic impact parameter dependence, indicating that overall laser absorption, accumulated over entire laser propagation inside the jet, does not highlight the essence of localized microlensing. The striking effect of curvature-driven forward microlensing in the cylindrical jet can be seen in the reflectivity signal coming from the first surface of the water jet, inset of Fig. 3A. The curvature-driven field enhancement occurs only progressively inside the liquid jet, and therefore leaves the input irradiance level at first jet-air interface unchanged, Fig. 2A. Consequently, the triggering of non-Fresnel response in the reflectivity mode requires nearly 9 times higher laser energy of 10.6 $\mu$J, than merely 1.1 $\mu$J needed in the transmission, Fig. 3A.    

\begin{figure}[h!] 
	\centering
	\includegraphics[width=1.0\textwidth]{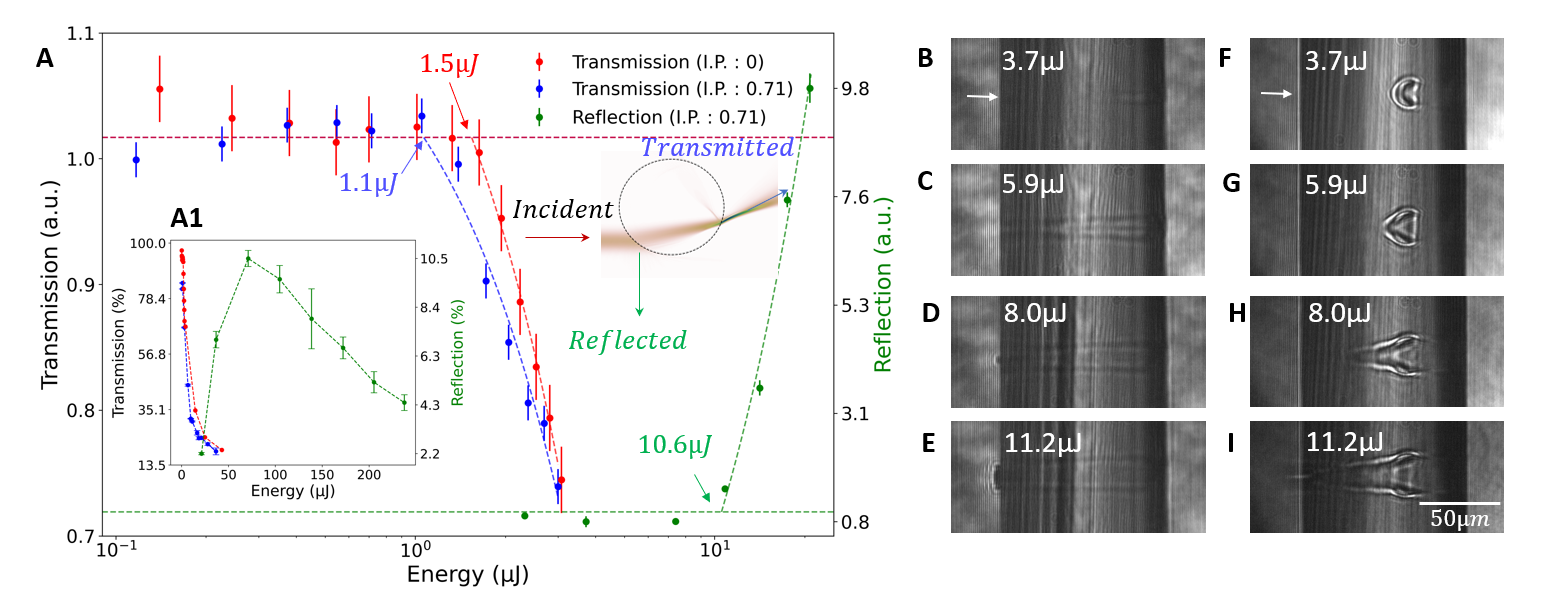} 
	\caption{\textbf{Dynamics of nonlinear absorption and plasma channels.}
 (\textbf{A}) Measurement of transmitted laser energy for symmetric and asymmetric cases, showing a transition from Fresnel to the non-Fresnel response of water jet. The reflectivity measurement for asymmetric geometry is also plotted in (A), with a common horizontal axis. The inset sketch in (A) shows the geometry of the incident, reflected and transmitted laser pulse under the asymmetric lensing. (\textbf{A1}) Transmission and reflection measurements similar to (A), but carried over a large laser energy range. Optical shadowgrams of laser energy-dependent evolution of the plasma channel, captured at a delay of 2ns, for symmetric (\textbf{B}) - (\textbf{E}) and asymmetric case (\textbf{F}) - (\textbf{I}). The white arrow in (B) and (F) indicates the laser propagation direction.}
\label{Conical Plasma Channel}
\end{figure}

The recorded optical shadowgrams provide further insights into the dynamics of input laser energy-dependent localized absorption, followed by the formation of plasma channels \cite{plasmawater2002, plasmawater2006, plasmawater2016, Dropshaping2015, plasmawater2018, plasmawater2020}. In both symmetric and asymmetric geometries, optical breakdown initiates near the rear surface of the water jet (Fig. 3B and F). However, the plasma channel contrast seen in shadowgrams is much more pronounced for the stronger asymmetric geometry. For input energy $\geq$ 5.9 $\mu$J, the symmetric case develops a faint, parallel-looking plasma channel, running from the front to the rear side of the jet, Fig. 3 C to E. In contrast, for the asymmetric case, the plasma channels evolve from a semi-spherical shape (Fig. 3F) to a conical shape (Fig. 3I). The conical channels are the manifestation of a much steeper enhancement of input laser irradiance (Fig. 2A), which in turn drives highly non-uniform laser absorption, followed by spatially varying plasma expansion.      

\begin{figure}[h!] 
	\centering
	\includegraphics[width=1.0\textwidth]{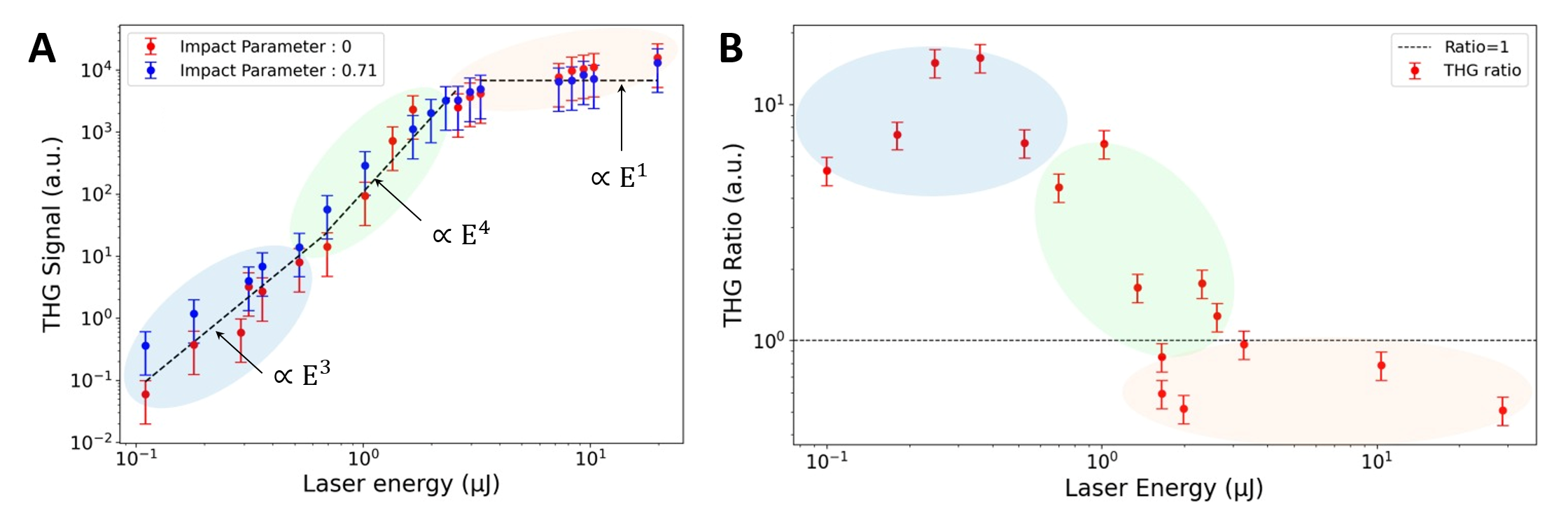} 
	\caption{\textbf{Microlensing driven phase-dynamics of THG signals.}
 (\textbf{A}) The variation of THG signal with input laser energy related to the symmetric (impact parameter = 0) and asymmetric (impact parameter = 0.71) geometries. (\textbf{B}) ratio of THG signal recorded for these two impact parameters as a function of laser energy.}
\label{THG microlensing}
\end{figure}

Apart from amplitude-sensitive diagnostics such as reflection, transmission (Fig. 3A) or optical shadowgrams (Fig. 3 B to I), we further investigated the dynamics of microlensing with a phase-sensitive process like THG \cite{BoydBook}. The plasma formation, driven by tight focusing of the incident field inside the water jet, can induce local, transient modifications of the refractive index and third-order nonlinearity, which in turn can affect the THG process due to phase matching requirements \cite{2009JETPDephasing}. The experiments, performed at low energy (Fig. 2B and C), were extended over a large energy range to capture multiple transition phases related to microlensing-driven plasma formation and subsequent impact on the THG signals. Basically, third-harmonic generation in an isotropic medium is affected by the presence of ionized, free electrons, which, by breaking symmetry, enable THG generation in a bulk medium. However, the presence of strong electron density introduces dephasing and phase mismatch, causing THG efficiency to drop \cite{2009JETPDephasing}. For symmetric and asymmetric cases, three distinct regimes could be identified where THG yield displays different laser energy ($E$) dependent power laws, Fig. 4A. The low energy region (0.1$\mu$J - 1.0 $\mu$J), where the THG signals are mostly originated from the liquid-air interface \cite{1995PRATHG}, shows $E^3$ dependence, Fig. 4A, highlighted by blue shade. In this range, both the reflection and transmission data show the Fresnel response (Fig. 3A) and no plasma channelling was observed in the shadowgram, indicating that THG signals are coming from the unionized interfaces. For the mid-energy range (1.0 $\mu$J - 3.0 $\mu$J), highlighted in green, the power law gets steeper ($E^4$), as the partial ionization in the bulk and at the boundary starts contributing towards the THG signal \cite{2009JETPDephasing}. Towards the high-energy ($>$ 3.0 $\mu$J), the THG yield crashes and slopes become slower than $E^1$. In this regime, the plasma channels are observed in both symmetric and asymmetric cases (Fig. 3C to E and 3G to I), which severely degrades the THG phase matching process \cite{2009JETPDephasing, 2011THGdephasing}. At higher laser energy ($>$ 5.0 $\mu$J), the measured THG signal may also include contributions from additional nonlinear effects such as four-wave mixing and self-phase modulation \cite{2002PRLDroplet, 2014JCPLiqidJet}. The onset of significant contributions from other nonlinear processes present at higher input laser energies was additionally verified with spectral and spatial observations (Fig. S5, S6).  

Besides individually examining the different phase dynamics of THG process, we further estimated the ratio of THG signal recorded under asymmetric and symmetric irradiation, Fig. 4B. In continuation of Fig. 4A, three distinct regimes can also be identified in the THG ratio, Fig. 4B. At the low energy, the strong microlensing assisted asymmetric geometry favours the field enhancement at the liquid-air interface and therefore THG ratio are nearly factor of 10. In the high energy range (red shaded area), the THG yield ratio flipped and the value drops below unity while going through a transition phase (Green area) in the mid-range. These results again highlight the dynamic role of microlensing in governing the outcome of initially interface-sensitive processes, followed by ionization-driven bulk response.

\subsection*{High-energy-density regime: caustic driven shocks in flowing microjets}
Laser-driven microexplosions offers pathways to access terapascal pressures in solid \cite{shock1984PRL, HighNASapphire, HighNATeraPascal} and liquid targets \cite{2016NatureXFEL, 2020PhysFluidNanosecond, 2022PhysFluidNanosecond}. To ensure localised energy absorption, previous experiments have generally relied on external, high-NA optics and have been performed mostly as single shots \cite{shock1984PRL, HighNATeraPascal} or at 10 - 100 Hz repetition rates \cite{2016NatureXFEL, 2020PhysFluidNanosecond, 2022PhysFluidNanosecond}. Here, in contrast, we employ caustic-driven microlensing, where the focusing element and interaction zones are realised in the same medium of a self-refreshing liquid jet. This configuration enables experiments at an unprecedented repetition rate of 0.2 MHz. Tight focusing achieved through asymmetric caustics lensing (Fig. 1E and F1) enables high-energy-density conditions, which in turn can trigger shock waves within the water microjets \cite{shock1996Vogel, shock1999Vogel}. As expected from the limited lensing capability offered by symmetric geometry, achieving high-energy-density conditions is less feasible than in the caustic-enhanced case. For instance, at a similar energy of 1.8 $\mu$J, the liquid jet under symmetric lensing seems entirely unperturbed by the laser interaction (Fig. 5A). Only on the longer time delay of 16.6 ns, a tiny plasma blob is detected at the liquid jet rear-side (Fig. 5B). After increasing the laser energy by nearly 7 times, a linear plasma channel is observed, which drives a faint cylindrical shock wave, propagating up and downstream of the liquid jet (Fig. 5E).  In the caustic geometry, a nearly spherical shock wave can be triggered with the laser energy as low as 1.8 $\mu$J (Fig. 5C). The bright localized plasma emission (Fig. 5C) coming from the rear side of the liquid jet corroborates the confined nonlinear absorption of laser energy. At the time delay of 7.4 ns, the shock front has clearly separated from the nearly stagnated plasma channel and carries the forward impact of the laser interaction in a far field. On later time scales of 10.9 ns(Fig. 5D), the spherical shock front continues to expand, while the shock wave seems to get distorted due to cylindrical lensing artifact caused by the liquid jet. The time evolution of the asymmetric lensing-driven shock front is shown in Fig. 5F, where for the maximum recorded time delay of 17 ns, the disturbance remains supersonic. Interestingly, at early time delays, the measured shock velocity of $\sim$ 3000 m/s corresponds to the shock pressure ($p_{shock}$) of nearly 2.3 gigapascal (see Methods). 

\begin{figure}[h!]
\includegraphics[width=0.8\textwidth]{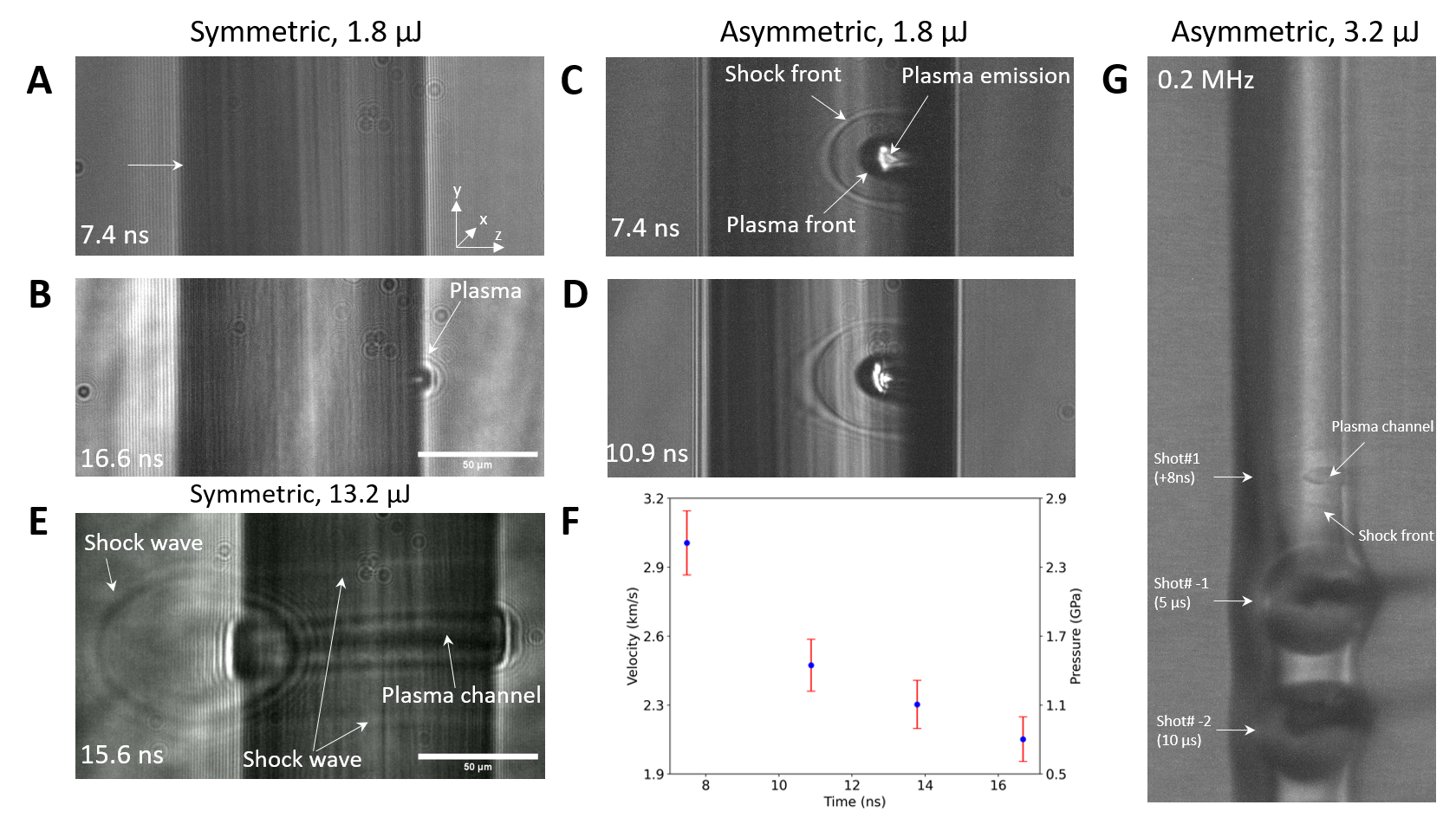}
\centering
\label{figure:5}
\caption{\textbf{Asymmetric lensing assisted shocks in microjets.}
\textbf{(A)} and \textbf{(B)} Shadowgraphic images of the laser interacting with the water jet under symmetric irradiation (impact parameter = 0) with pulse energy of 1.8  $\mu$J, at a time delay of 7.4 ns and 16.6 ns, respectively. \textbf{(C)} and \textbf{(D)} Shadowgraphic images under the asymmetric geometry (impact parameter = 0.71) with pulse energy of 1.8  $\mu$J, captured at a time delay of 7.4 ns and 10.9 ns, respectively. \textbf{(E)} Linear plasma channel and faint shock waves formation under symmetric irradiation after increasing laser energy to 13.2  $\mu$J at a time delay of 15.6 ns. \textbf{(F)} The velocity and estimated pressure of shock waves inside the liquid jet corresponding to the asymmetric case shown in (C) and (D). \textbf{(G)} Driving microexplosions at a repetition rate of 0.2 MHz under the asymmetric case, with input energy of 3.2 $\mu$J. The previous back-to-back two shots, each delayed by 5$\mu$s, due to 0.2MHz repetition rate, are also indicated.}
\label{Shadow Image}
\end{figure}

For the symmetric excitation, the weaker extended plasma channel lacks the energy density required to launch a strong shock wave, unless assisted by ramping up the external laser energy, all the way up to 13.2$\mu$J, which surpassed the triggering of ionization at the front surface, as observed in non-fresnel reflectivity measurements occurring of energy $>$ 10$\mu$J. Surpassing the front surface plasma formation thresholding also drives a strong shock wave in the air-liquid interface. In contrast, the strongly localized energy deposition for asymmetric case facilitates the launching of the shock wave predominantly inside the liquid jet, without disturbing the front surface of the liquid jet. Moreover, at these low thresholds, the air breakdown at the jet front surface also can be avoided.

\section*{Discussion and Conclusion}
Our results show that caustic microlensing plays a broad role extending well beyond conventional geometrical focusing.  Here, the refractive lensing and light-matter interactions are realized in the very same medium of liquid jet, resulting in an optical system that is self-aligned and does not require any external alignment. The one-dimensional microlensing of the Gaussian beam in the cylindrical liquid jet, producing an Airy pattern, may also enable advanced imaging such as light-sheet microscopy \cite{2017Lightsheet}. In contrast to the high-NA objectives \cite{1997MazurAPL, HighNATeraPascal, HighNAVoid2006, VogelPRL2008, HighNASapphire}, which must be placed close to the target and are easily degraded by ablation debris, our caustics lensing requires deployment of a low-NA input optics positioned far from the interaction zone and thereby enables long-term, debris-free operation. Furthermore, the continuous flow of the liquid jet ensures renewal of the irradiated optical surface after each pulse, which facilitates reproducibility and thermal stability even at 0.2 MHz repetition rate. This self-refreshing and alignment-free geometry provides an experimental platform to drive high-intensity ultrafast applications where conventional focusing may struggle. 

To enhance the nonlinear signals, the caustic microlensing intrinsically localizes the focused light field at the liquid–air interface. This self-aligned localization boost the surface-sensitive nonlinear responses, such as Sum frequency generation (SFG), difference frequency generation (DFG) and third-harmonic generation (THG), without requiring micrometer-scale alignment by high-NA external focusing optics. It also turns out that these interface locatization are quite stable against variations in the liquid jet diameter, and therefore could immensely be beneficial in the field of interface-sensitive nonlinear spectroscopy.

Moving further in the high-intensity domain, the confinement of focused light intensity by caustic microlensing suppresses extended volumetric ionization and plasma channels. These geometric localization enables the formation of compact micro-plasma structures and on nanosecond time-scales, driven giga-pascal scale shock waves using only microjoule-level femtosecond pulses. Our results may offer exciting opportunities to pursue high-energy-density (HED) science at MHz class repetition rates. While terapascal-level pressures have been reported in solid dielectrics under extreme focusing achieved by external high-NA optics \cite{HighNATeraPascal, HighNAVoid2006, HighNASapphire}. However, those approaches are based on static configurations, require precise alignment, and are not readily scalable to high-repetition rates. In contrast, the present work demonstrates that comparable spatial confinement and energy densities can emerge naturally from caustics within a dynamically refracting, self-renewing liquid medium.

In conclusion, our results establish caustic-driven microlensing in a self-refreshing cylindrical liquid jet as a simple and scalable route to extreme light localization and high-repetition-rate laser–matter interactions. The effectiveness of caustic lensing spans regimes from linear optics (surface-localized focusing) to nonlinear optics (enhanced third-harmonic generation) and high-energy-density physics (GPa-scale microshocks). The self-lensing jet provides a tunable, debris-free, and thermally stable optical element that simultaneously serves as target and lens, opening new possibilities for surface-specific nonlinear optics, high-throughput ultrafast spectroscopy, and table-top studies of warm dense matter and shock phenomena. Looking ahead, integrating such caustic-microlensing platforms with thin-disk mode-locked oscillators, delivering microjoule-level femtosecond pulses at multi-MHz rates\cite{2008NatPhotOscillator}, offers exciting prospects for optical pump–X-ray probe studies at XFELs \cite{2014Chapman, 2021NatureXFEL, 2016NatureXFEL}, and enables the generation of large, high-quality datasets ideal for benchmarking AI and machine-learning models of matter under extreme conditions\cite{2021NatureAIdata}.


\clearpage 

%

%
%
%
%
%
%


\section*{Acknowledgments}
\paragraph*{Funding:}
The authors acknowledge support of the Department of Atomic Energy, Government of India, under Project Identification No. RTI4007.

\paragraph*{Data availability:}
All data needed for evaluations of conclusions in the paper are mentioned in the main and/or in the Supplementary sections. Any additional data related to this paper can be requested from the authors.


\newpage


\renewcommand{\thefigure}{S\arabic{figure}}
\renewcommand{\thetable}{S\arabic{table}}
\renewcommand{\theequation}{S\arabic{equation}}
\renewcommand{\thepage}{S\arabic{page}}
\setcounter{figure}{0}
\setcounter{table}{0}
\setcounter{equation}{0}
\setcounter{page}{1} 


\begin{center}
\section*{Supplementary Materials: \scititle}

        Sourabh~Singh$^{1}$,    
	S.~Sree~Harsha$^{1}$,
        Tamanna$^{1}$,
	Prashant~Kumar~Singh$^{1\ast}$\and
	
    \small$^{1}$Tata Institute of Fundamental Research Hyderabad, 36/P Gopanpally, Hyderabad 500046, Telangana, India \and
	\small$^\ast$Corresponding author. Email: pks@tifrh.res.in
\end{center}




\subsection*{Supplementary Text}

\subsection*{Experiment details}

\subsubsection*{Femtosecond Laser system for driving micro-explosions}
The experiments reported in this paper were carried out at TIFR-Hyderabad India. The liquid jets were irradiated with a commercial Yb:KGW laser (Pharos, Light Conversion Ltd.) providing 2000 $\mu$J pulse energy, central wavelength of 1034 nm, pulse duration of about 170 fs and maximum repetition rate of 200 kHz. The details of the experiment set-up is sketched in Figure S3A. The laser beam of 7 mm diameter, propagating horizontally, and perpendicular to the cylindrical liquid jet axis, was focused  with a gold-coated off-axis parabolic mirror (f = 100 mm). The laser focal spot, having beam waist ($\omega_{o}$) distribution of 11.4 $\mu$m  $\times $ 12.9 $\mu$m (Figure 1C), was measured by 20$\times$ Mitutoyo NIR Objective, having a numerical aperture of 0.42. Finally, with 170 fs pulse duration, 1.0 µJ of energy, and the above focal spot distribution, the estimated peak intensity during the interaction was $1.27\times 10^{12} W/cm^2$.

\subsubsection*{Formation and characterization of laminar liquid jet}
 All the experiments with liquid jets were carried out at ambient temperature ($25^\circ C$) and atmospheric pressure. Deionized water was selected as a choice of liquid, however, results were qualitatively similar with methanol as well. A liquid container was pressurized at a backing pressure of 2 bar which resulted in the downward flowing laminar water jet through a home-built metallic nozzle fabricated by laser micromachining. Beyond the laminar regime, the liquid jet surface was subjected to the Rayleigh-Pleatue perturbations, which eventually leads to the break up of the jet into a random sequence of droplets. Therefore, to ensure smooth lensing from the unperturbed liquid jet,  the laser interaction was kept just 2 to 3 mm below the nozzle opening.  The diameter of the liquid jet was in the range of 100 $\mu$m - 110 $\mu$m.  During the experiment, the position of the laser interaction with the liquid jet was controlled by a manual three-axis translation stage. 

The water jet velocity was maintained between 10  m/s and 20 m/s depending on the experiment, corresponding to a displacement of 100 $\mu$m between successive pulses at 0.2 MHz repetition rate. This displacement exceeds the spatial extent of the plasma, cavitation, and shock region produced by a single pulse, ensuring that each interaction probes an undisturbed volume of liquid. To observe the effects of three to four successive laser shots, a lower magnification was used to capture a larger field of view (Fig. 5G). This led to a change in the imaging resolution and visual appearance of the images. Despite this, the consistent shock speed measurements confirm that the effects of previous pulses were successfully avoided.

\subsubsection*{Selective irradiation at different impact parameter}
For the impact parameter of 0, due to symmetric geometry, the laser pulse after refracting from the front and rear surface of the liquid jet would continue to follow the original path traced by the laser pulse in the absence of the liquid jet. This idea was implemented for alignment of the liquid jet under the symmetric geometry. The Fresnel reflection from the front surface also propagates backwards to the OAP, however, this monitoring protocol was less followed. For an impact parameter of 0.71, the geometric reflection from the first Fresnel reflection makes a 90-degree angle. The reflected beam pointing was monitored on a screen which was imaged on a CMOS camera. The highest impact parameter tried to end up having a reflection beam making an angle of 100 degrees; however, at this geometry, being at the grazing angle, a part of the laser beam also starts missing the liquid jet.

\subsubsection*{Pump-probe imaging for plasma channels and shock}
The space and time-resolved shadowgraphic images of the jet-laser interaction were captured on an 8-bit CMOS camera. A time-delayed probe beam, split from the main pulse, was frequency doubled in a  BBO crystal. The optical probe delay line was extended in the nanoseconds regime to capture the long-lasting dynamics of the shock wave (Fig. 5). For uniform background imaging, diffusers were used to make the probe beam spatially incoherent. To avoid fundamental pump scattering and plasma emission dominating the shadowgrams, the BG39 filter was used. 

\subsubsection*{Third Harmonic Generation in liquid jet}
The Third Harmonic Generation (THG) from the liquid jet laser interaction was measured in the forward transmission direction. The THG signal can be generated from the laser pulse focused in air. However, under our parameter range, the contribution from the air-plasma driven THG was measured to be negligible compared to what originated from the liquid-jet laser interaction. The THG signal, generated from the liquid jet (Fig. 2, B and C) was imaged by the same 20$\times$ Mitutoyo NIR Objective, with a BG-39 filter to select the third-harmonic signal. The THG yield (Fig. 4), was measured by imaging the far field profile of the third harmonic on a screen, using a CMOS camera protected with a BG-39 filter. 

\subsubsection*{Reflection and Transmission measurement of liquid-jet laser interaction}
The reflection and transmission measurements from the laser-jet interaction was performed with two types of power meter: OPHIR (3A-P-QUAD, power range: 100 $\mu$W - 3W) and Thorlabs (S310C, power range: 10 mW - 10 W). We use the OPHIR detector to measure the reflection from the liquid jet (Fig. 3, A and A1), which is sensitive to low-power measurements. As reflectivity involves participation of only the front surface where the volumetric lensing phenomenon is absent, the reflection measurements were performed under asymmetric irradiation. The transmission measurement (Fig. 3, A and A1) was performed for both symmetric and asymmetric geometries with the ThorLab detector, having a larger energy range.   

\subsubsection*{Numerical Simulation}
The numerical simulation to capture the lensing features in dielectric cylindrical liquid jet was carried out using the Wave Propagation Model (WPM) module of Diffractio \cite{2024diffractio}, an open-source NumPy-based Python module. Diffractio uses Rayleigh–Sommerfeld diffraction integral \cite{Fabin2006} to determine the wave propagation from the source, which deals with the wavefront of electromagnetic waves. Parameters, similar to the experiment, such as laser wavelength (1034 nm), dielectric refractive index of water (1.33) and liquid jet diameter (100 $\mu$m) were used in the simulation.  The simulation box, having 200 $\mu$m space, was subdivided into 3000 grids, resulting in a spatial resolution of 0.067 $\mu$m, which is much smaller than the wavelength of the laser. Scalar mask XZ function from Diffractio was used to modify the optical wavefront by blocking, transmitting, or phase-shifting different regions of the wave. Additionally, the Scalar source X function in Diffractio was used to define the initial wavefront, such as a plane wave, Gaussian beam, or other waveforms, which then propagates and interacts with several optical elements like lenses, apertures, or masks.  The numerical results presented here were also independently verified by using COMSOL Multiphysics. 

\subsubsection*{Calculation of Shock pressure}
The strength of the shock pressure generated during the propagation of shock wave can be estimated from the Rankine-Hugoniot conditions, which describe the conservation of mass, momentum, and energy across a shock front. For shock save speed of $v_{shock} = 3000 m/sec$, ambient mass density of water, $\rho_{water} = 1000 kg/m^3$, and sound speed in water, $c_{sound} = 1500 m/sec$, the shock pressure is given as $P_{shock} = \rho_{water}\times c_{sound}\times (v_{shock} - c_{sound})$ = 2.25 gigapascal \cite{1948Cole}.

\subsubsection*{Caustics driven finite-energy Airy pattern under asymmetric irradiation}

\begin{figure}[h!] 
	\centering
	\includegraphics[width=0.75\textwidth]{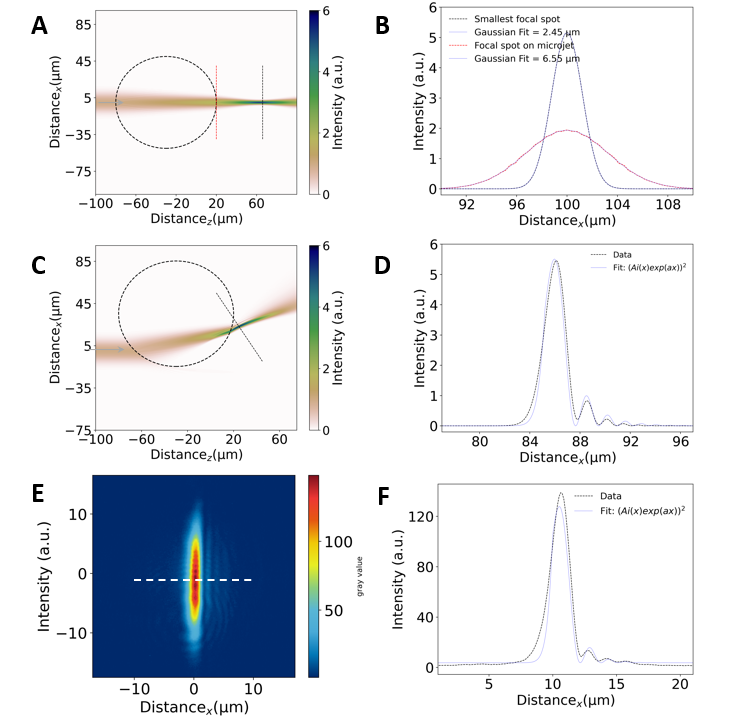} 

	\caption{\textbf{Finite-energy Airy pattern in Asymmetric lensing}
		\textbf{A}) Simulation results for dielectric lensing under symmetric irradiation, where focusing occurs outside the liquid jet. \textbf{B}) The focal spot profile at the rear surface of the jet and the smallest focal spot occurring nearly 40$\mu$m further downstream. \textbf{C}) Simulation for asymmetric lensing where focusing occurs at the rear surface of the liquid jet. \textbf{D}) The lineout of the focal spot has asymmetric wings, which are fitted with an exponentially decaying Airy function. \textbf{E}) Experimentally measured beam profile at impact parameter of 0.71. \textbf{F}) Measured intensity variation also shows exponential decaying Airy behaviour.}
	\label{fig:sup_example} 
\end{figure}

Figure S1(A) depicts the numerical simulation results for the symmetric irradiation (impact parameter = 0). Under this geometry, the relaxed microlensing conditions cause the smallest focal waist to be realized outside the circle (Fig. S1A). Two vertical line cuts (red and black, Fig.S1A) are taken to obtain the beam-waist distribution at the surface of the liquid jet and at the smallest focal plane, respectively, which turns out to be 6.6 $\mu$m and 2.5 $\mu$m (Fig. S1B)). In contrast, for the asymmetric geometry (impact parameter = 0.71), the strong microlensing ensures that the tightest focusing plane occurs at the rear surface of the liquid jet. Moreover, due caustics effect, the beam profile at the smallest focal plane displays non-symmetric wings (Fig. S1C). These distributions are fitted with the exponentially decaying Airy function (Fig. S1D), which represents the features of finite-energy Airy beams \cite{2007OLFiniteAiry}. Corroborating with the simulation, the experimentally measured fundamea characteristic of fold catastrophe, which can also be modelled by an exponentially decaying Airy function (Fig. S1F). 

\begin{figure}[h!] 
     \begin{center}
     \includegraphics[width=0.4\textwidth]{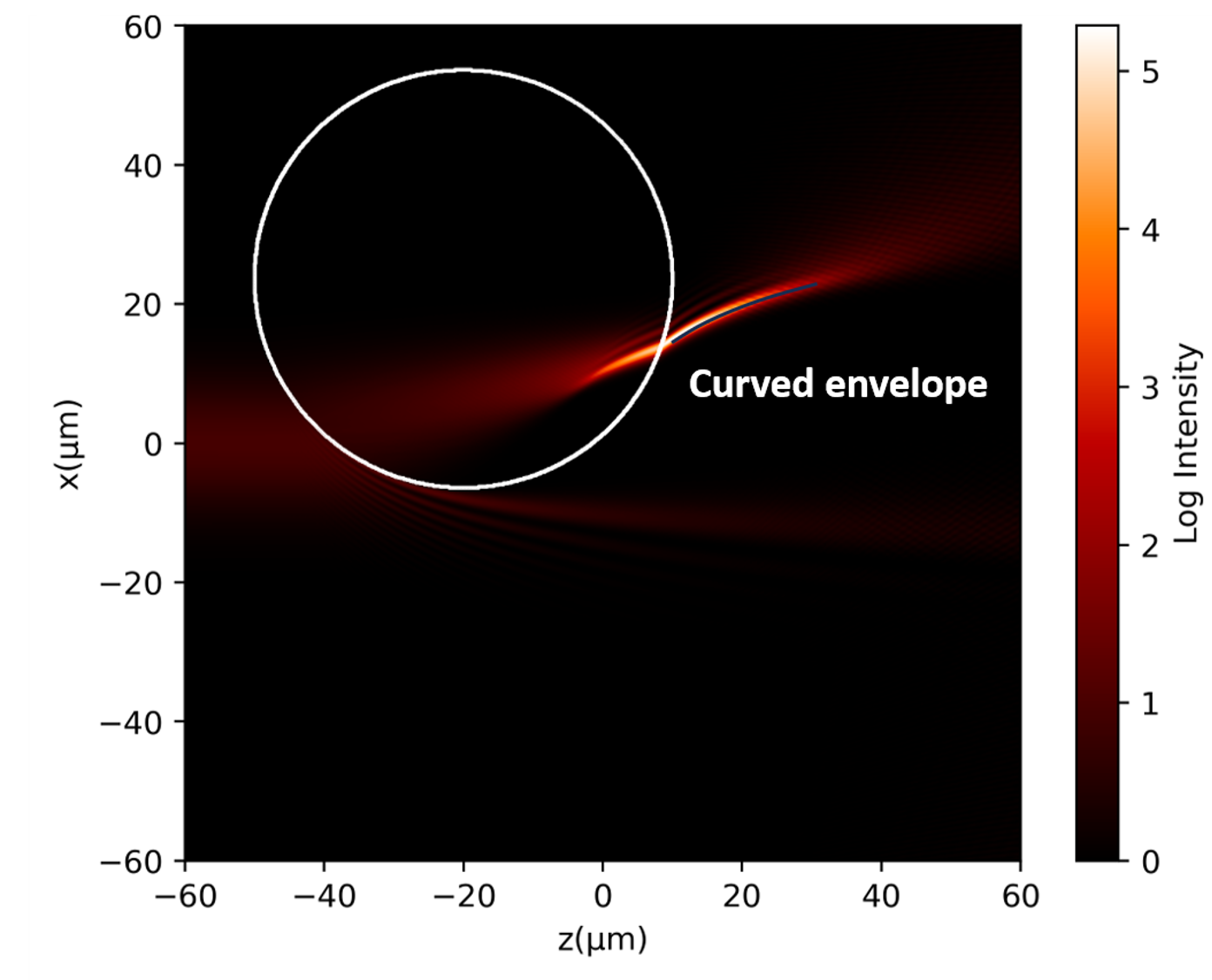}
     \caption{Caustics-assisted microlensing producing a curved trajectory.}
     \end{center}
\end{figure}

\begin{figure}[h!] 
     \begin{center}
     \includegraphics[width=1\textwidth]{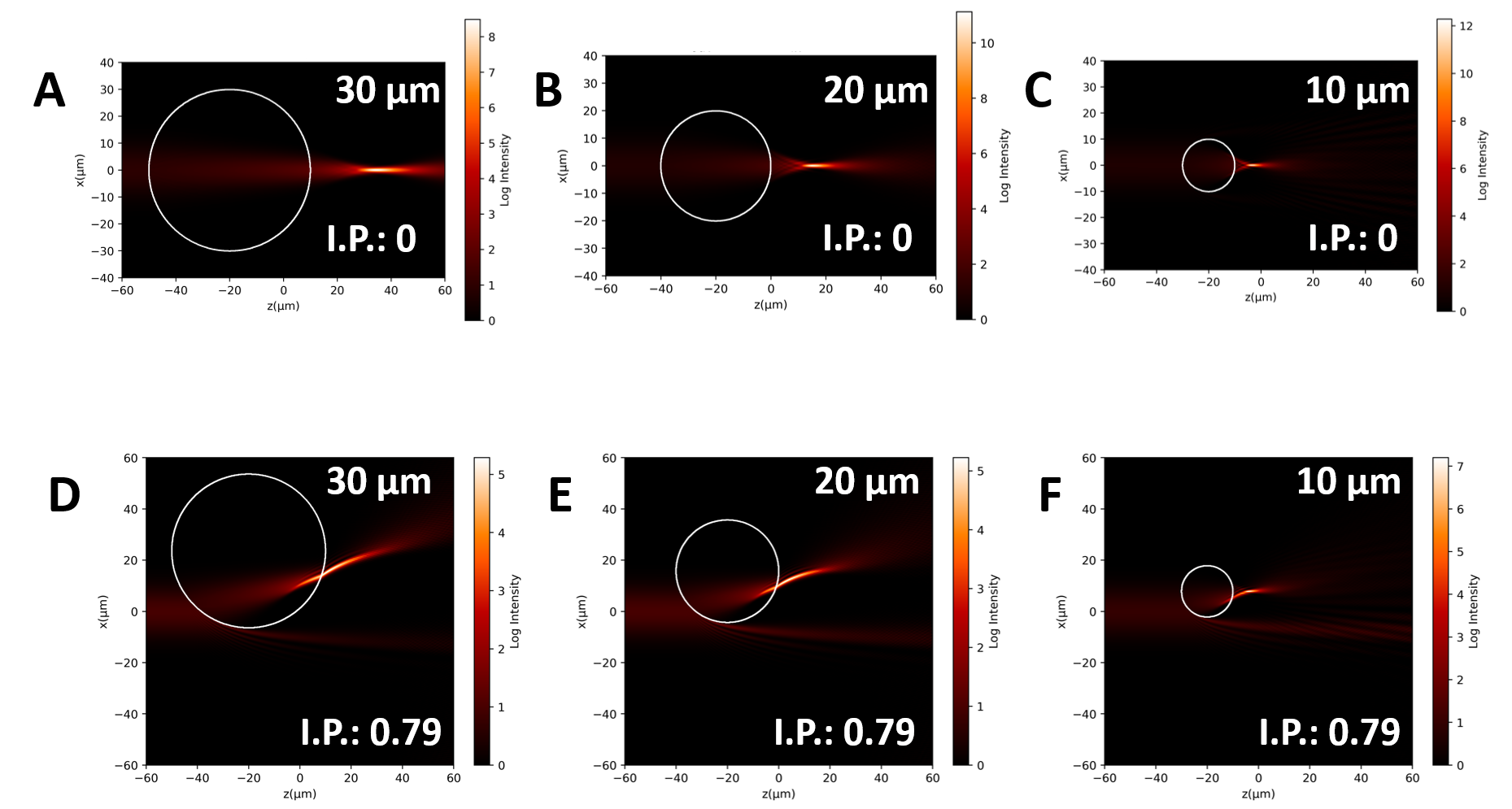}
     \caption{ Microlensing simulations for symmetric (A - C) and asymmetric (D - F) geometry by varying the radius of curvature of liquid droplets.}
     \end{center}
\end{figure}

In contrast to the symmetric lensing, where focusing of input light occurs along a straight line (Fig. S1A), the caustics microlensing driven locus of high-intensity point traces a curved path (Fig. S2). This curved trajectory is a classical signature of caustics and is somewhat stable against changes in the lensing diameter. For example, when the liquid-jet diameter is reduced from 30 µm to 10 µm, the region of maximum intensity consistently resides at the liquid–air interface (Fig. S3D–S3F). This behaviour contrasts sharply with that observed under symmetric or external lensing, where changes in jet size or focal length can noticeably shift the location of the most intense region (Fig. S3A–S3C).



\begin{figure}[h!] 
	\centering
	\includegraphics[width=0.8\textwidth]{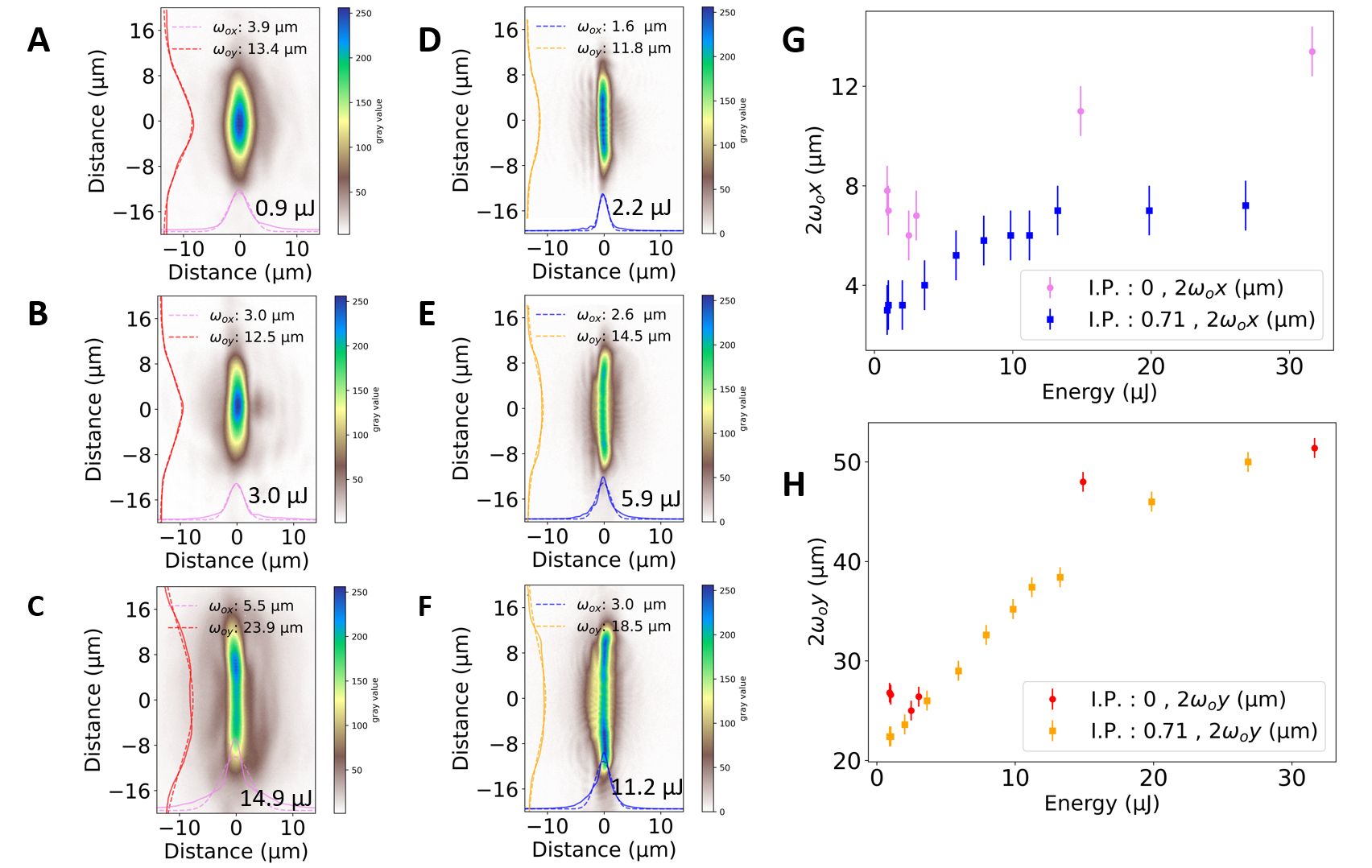} %

	\caption{\textbf{Plasma effects on liquid-jet driven microlensing}
		(\textbf{A}), (\textbf{B}) and (\textbf{C}) show the effect of increasing the input laser energy on the lensing realised under the symmetric irradiation (impact parameter =0) of the liquid jet. (\textbf{D}), (\textbf{E}) and (\textbf{F}) represents these effects under the asymmetric irradiation (impact parameter =0.71). (\textbf{G}) and (\textbf{H}) shows the input energy dependent behaviour of the focal spot along (2$\omega_{0x}$) and perpendicular (2$\omega_{0y}$) to the cylindrical lensing axis, under impact parameters of 0 and 0.71.}
	\label{fig:sup_example} 
\end{figure}

\subsubsection*{Plasma effects on liquid-jet driven microlensing}

In the linear regime, the micro-lensing occurring due to the dielectric cylinder is governed by the impact parameter (Fig. 1D and E). However, in the non-linear regime, the refractive index gets strongly modified and in the event of optical breakdown of the liquid jet, the plasma medium starts dominating over the linear lensing (Fig. 3). To capture transitions from linear to non-linear regime, the fundamental focal spot was monitored by systematically increasing the input laser energy, Fig. S4. Under the symmetric irradiation (Fig. S4 A, B, G and H), the focal waist ($\omega_{0x}$ and $\omega_{0y}$) initially remains constant in the low energy range ($\sim 3 \mu$J), followed by an increase in the beam waist (Fig S4 C, G and H) for higher laser energy. Similar observations were recorded for the asymmetric geometry (Fig. S4 D to H). However, the difference comes from the fact that, despite plasma effects, the focal waist along the cylinder lensing axis (2$\omega_{0x}$) for impact parameter 0.71, remains less compared to the focal spot corresponding to impact parameter 0 (Fig. S4 G). As expected, the focal waist perpendicular to the cylinder lensing axis (2$\omega_{0y}$) does not show any dependence on the impact parameter (Fig. S4 H). 


\begin{figure}[h!]
\centering
\includegraphics[width=\textwidth]{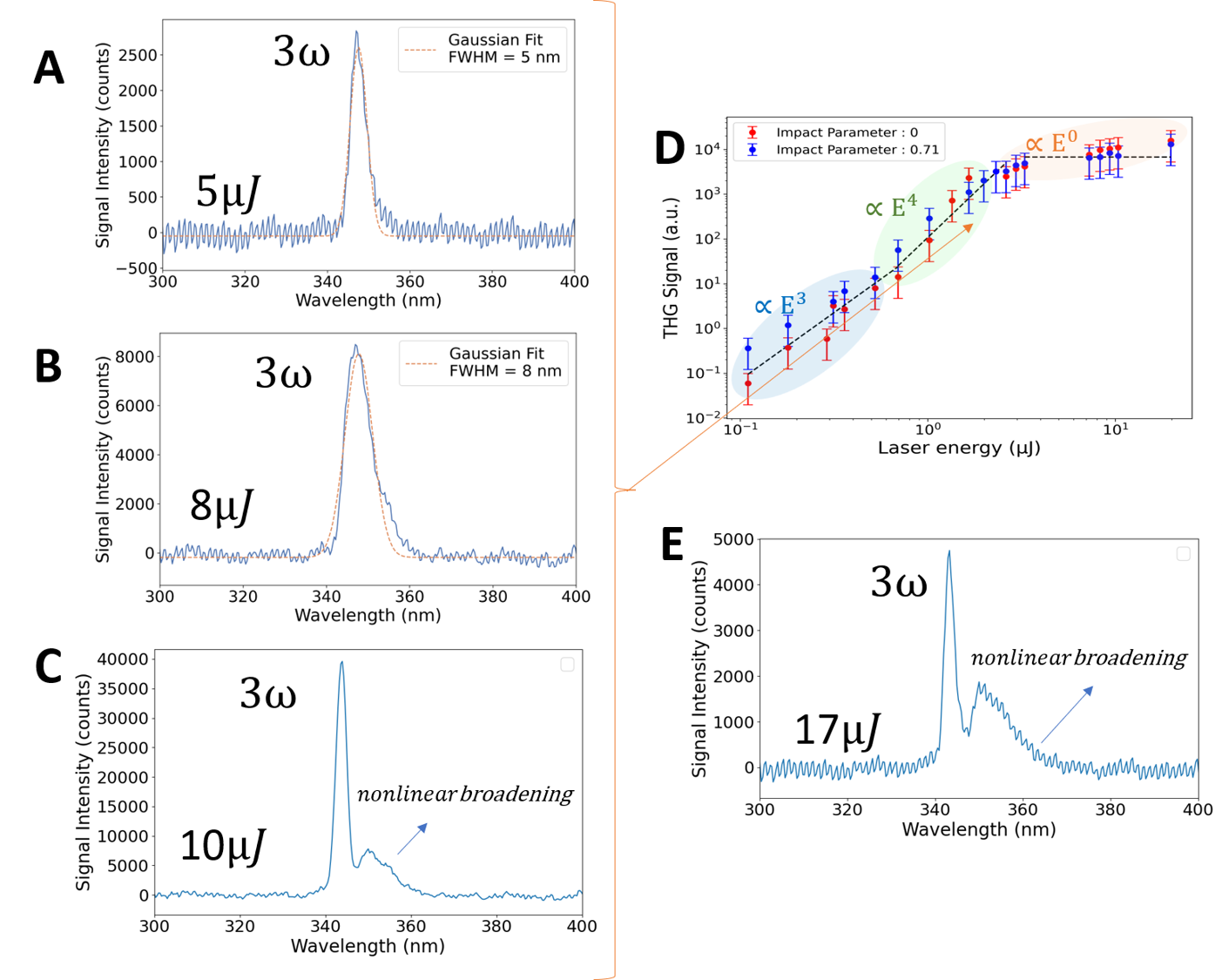}

\renewcommand{\thefigure}{S\arabic{figure}} 
	\caption{\textbf{Third harmonic optical spectrum} (\textbf{A}), (\textbf{B}), (\textbf{C}) and (\textbf{E}) shows THG spectrum at impact parameter of 0, laser energy of 5  $\mu$J,  8  $\mu$J, 10 $\mu$J and 17 $\mu$J. (\textbf{D}) The variation of THG signal with input laser energy related to the symmetric (impact parameter = 0) and asymmetric (impact parameter = 0.71) geometries.}
\label{Shadow_Image1}
\end{figure}

\begin{figure}[h!]
\centering

\renewcommand{\thefigure}{S\arabic{figure}} 

\includegraphics[width=\textwidth]{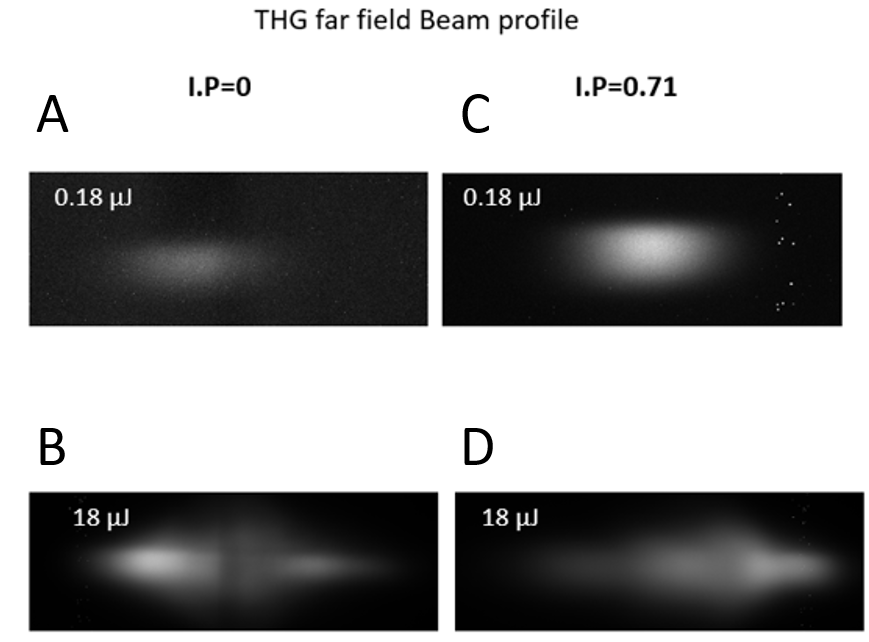}
	
	\caption{\textbf{Third harmonic Spatial response} Spatial response of nonlinearity in the measured THG beam far field profil. (\textbf{A}), (\textbf{B}), (\textbf{C}) and (\textbf{D}) show the far field beam profile at impact parameter of 0 and 0.71, at energies of 0.18  $\mu$J and 18 $\mu$J, respectively.}
\label{Shadow_Image2}
\end{figure}

\subsubsection*{Spectral and Spatial signatures of nonlinearities present in the generated THG beam}

At higher laser intensities, the measured THG signal may include contributions from additional nonlinear effects such as four-wave mixing and self-phase modulation. To probe these effects, we examined both the spectral (Fig. S5) and spatial (Fig. S6) characteristics of the generated THG beam. For input laser energies of 5$\mu$J - 8$\mu$J,  the THG spectrum (Fig. S5A,B) exhibits noticeable broadening, while at even higher energies ($> 10\mu$J, Fig. S5C, S5E ), an asymmetric spectral sideband emerges, deviating from the low-energy THG profile.

Similarly, in the spatial domain (Fig. S6), the smooth beam profile observed at laser energy of 0.18$\mu$J (Fig. S6A,S6C), becomes strongly distorted at 18$\mu$J (Fig. S6B, S6D). Together, these spectral and spatial observations confirm the onset of significant contributions from other nonlinear processes present at higher input laser energies.


\subsubsection*{Micro-explosions being independent of Laser polarization}

Except for the symmetric irradiation, under any non-zero impact parameter, the electric field of the incident laser pulse can have a component perpendicular to the cylindrical liquid jet axis (horizontal polarization) or be parallel to the jet axis (vertical polarization). At the front surface of the liquid jet, the choice of polarization can make a big difference in term of non-linear plasma coupling. However, as the micro-lensing process starts acting beyond the front interface (Fig. 2A), the formation of the plasma channel and micro-explosions should be independent of the input laser polarization. To substantiate these ideas, shadowgraph images were recorded for both horizontal and vertical polarization (Fig. S7). The basic sketch of the experimental setup is shown in Fig. S7A and described in Methods. As illustrated in the recorded shadowgrams (Fig. S7 B-I), the plasma formation seems to be independent of the laser polarization. Therefore, under these conditions, the impact parameter turns out to be a strong parameter determining the laser-plasma coupling (Fig. S7 D1, H1 and J).

\begin{figure}[h!] 
	\centering
	\includegraphics[width=1.0\textwidth]{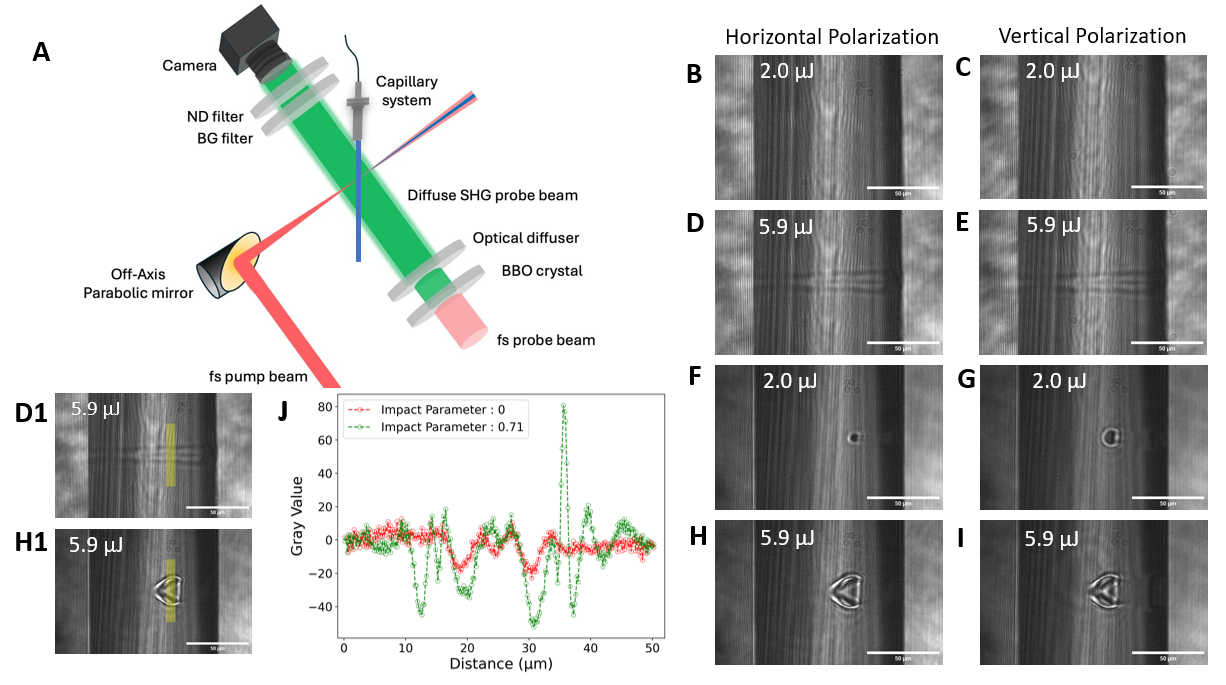} 

	\caption{\textbf{Polarization independent coupling}
	(\textbf{A}) displays a sketch of the experimental setup for pump-probe shadowgraphy. (\textbf{B}) and (\textbf{D}) are plasma channels under symmetric irradiation for horizontal laser polarization, whereas (\textbf{C}) and (\textbf{E}) are related to the vertical polarization. (\textbf{F}), (\textbf{G}), (\textbf{H}) and (\textbf{I}) correspond to similar conditions as (B - E), except done under asymmetric irradiation. (\textbf{J}) shows the linecut taken from \textbf{D1} (impact parameter  = 0) and \textbf{H1} (impact parameter = 0.71). All the above shadowgrams are taken at a temporal delay of 2 nanoseconds.}
	\label{fig:sup_example} 
\end{figure}


\subsubsection* {Possibilities of driving micro-explosions experiments towards MHz Repetition rate}

The asymmetric irradiation-driven microlensing enables the realisation of a gigapascal-scale pressure wave at input laser energy of just a few microjoules. Due to the localised high-energy-density conditions, the bulk of the liquid jet is not perturbed. On a longer time scale lasting several microseconds, the relaxation of the heated region occurs via physical ejection or ablation of particles. The ablated volume, which is largely governed by the absorbed laser energy, affects the structural integrity of the neighbourhood region of the liquid jet (Fig. S4C).

\begin{figure}[h!] 
	\centering
	\includegraphics[width=1.0\textwidth]{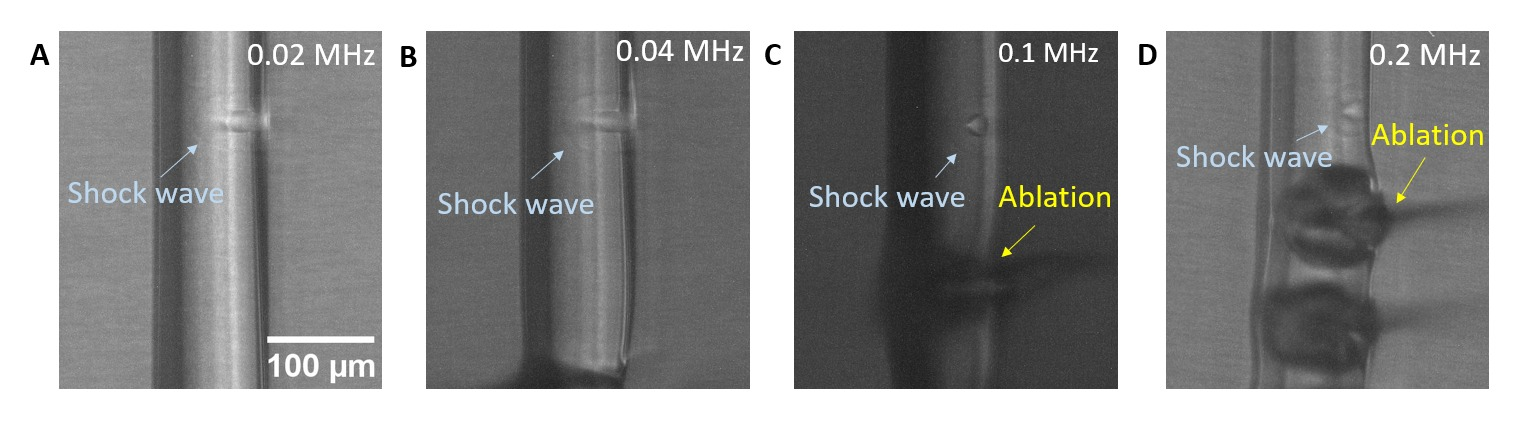} 

	\caption{\textbf{Scaling the repetition rate.}
	(\textbf{A}), (\textbf{B}), (\textbf{C}) and (\textbf{D}) formation shock waves, unaffected by post-shot ablation, at repetition rates of 0.02 MHz, 0.04 MHz, 0.1 MHz and 0.2 MHz, respectively. All these shock waves are captured at 12ns delay, with the impact parameter of 0.71 and laser energy of 3.2  $\mu$J. The liquid jet was flowing with a velocity of 11m/s (A and B), 15 m/s (C) and 20 m/s (D).}
	\label{fig:sup_example} 
\end{figure}

\begin{figure}[h!] 
	\centering
	\includegraphics[width=1.0\textwidth]{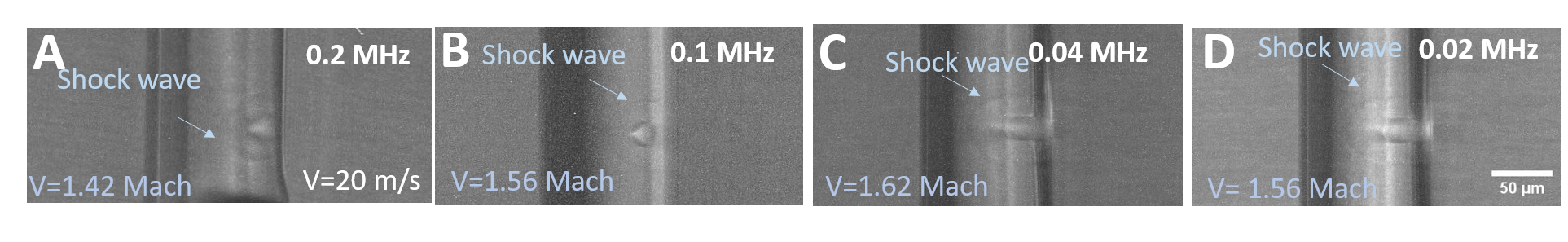} 

	\caption{\textbf{Zoomed version for: Scaling the repetition rate.}
	(\textbf{A}), (\textbf{B}), (\textbf{C}) and (\textbf{D}) formation shock waves, unaffected by post-shot ablation, at repetition rates of 0.2 MHz, 0.1 MHz, 0.04 MHz and 0.02 MHz, respectively. All these shock waves are captured at 12ns delay, with the impact parameter of 0.71 and laser energy of 3.2  $\mu$J. The liquid jet was flowing with a velocity of  20 m/s (A), 15 m/s (B), 11m/s (C and D).}
	\label{fig:sup_example} 
\end{figure}

On increasing the laser repetition rate (Fig. S8 A to D), the likelihood of the interaction area getting affected by the previously ablated shots increases. To avoid the possibility of multi-pulse interactions, where laser pulses hit the same spot on the jet, the gap between two consecutive laser shots can be increased by boosting the liquid jet speed. To exclude possible cumulative effects from successive pulses, we repeated the measurements at repetition rates reduced by an order of magnitude (0.02 MHz). The observed shock speeds were unchanged within experimental uncertainty, confirming that each event is governed by single-pulse dynamics rather than multi-pulse accumulation (Fig. S9). The invariance of the measured shock speed between 0.02 MHz and 0.2 MHz confirms the absence of cumulative or multi-pulse effects (Fig. S9 A to D). Following these ideas, we have succeeded in demonstrating non-overlapping interaction at a repetition rate of 0.2 MHz (Fig. S8D), with jet flowing at 20m/s.  A further scaling of the repetition rate should be possible by tuning the liquid jet velocity and diameter \cite{2018OptExpLiquidJet}.


\subsubsection* {Shaping the geometrical structure of the Shock Wave}

The geometrical shape of the shock wave is largely governed by the way the nonlinear absorption of laser energy takes place in the liquid jet. For asymmetric irradiation (impact parameter =0.71), at the threshold of plasma formation, the energy deposition occurs in a localised region (Fig. 2E), which drives a nearly spherical shock (Fig. S10A). On increasing the input laser energy, the ionization threshold conditions are met in the early phase of laser focusing inside the liquid jet. However, the presence of steep intensity gradient along the laser propagation (Fig. 2A) leads to non-uniform absorption of the laser energy, which drives a conical-shaped plasma channel (Fig. 3 H and I). On longer time delays, the shock wave after getting detached from this conical plasma channel retains the conical shape, as demonstrated in the shadowgraphic data (Fig. S10B). 

\begin{figure}[h!]
\centering

\renewcommand{\thefigure}{S\arabic{figure}} 

	\includegraphics[width=\textwidth]{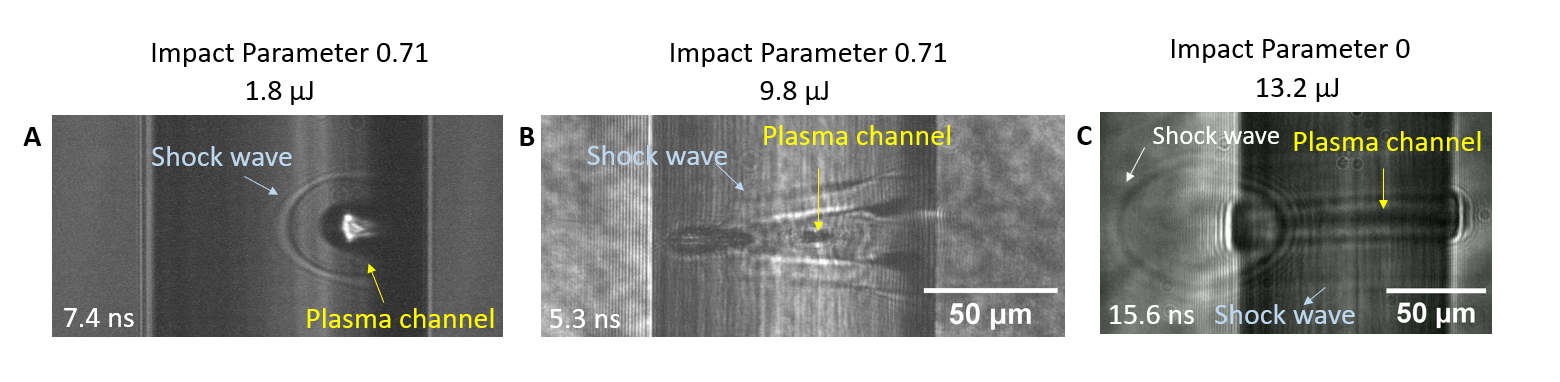}
	\label{figure 1: 50 shot electron beam distribution and stability:  }
	\caption{\textbf{Generating Shock Waves of different shapes} (\textbf{A}) Spherical shock at impact parameter of 0.71, laser energy of 1.8  $\mu$J and temporal delay of 7.4 ns. (\textbf{B}) Conical shock at impact parameter of 0.71, laser energy of 9.8 $\mu$J, delay of 5.3 ns. (\textbf{C}) Cylindrical shock in liquid jet at impact parameter of 0, laser energy of 13.2$\mu$J, and delay of 15.6 ns.}
\label{Shadow_Image3}
\end{figure}

For symmetric irradiation (impact parameter =0), as the cylinder curvature appears nearly flat, the laser intensity variation along the propagation direction shows relatively slower increasing trend (Fig. 2A), resulting in a cylindrical plasma channel (Fig. 3F). As a result, the surrounding plasma expands cylindrically, giving rise to cylindrical shock waves (Fig. S10C), inside the liquid jet. Additionally, a spherical shock wave, expanding in the air, is also observed at the front of the liquid jet, attributed to localized heating of the front surface (Fig. S10C). In summary, by taking advantage of the local curvature-dependent micro-lensing occurring in the liquid jet, one could drive shock waves of different geometrical shapes.



\end{document}